\documentclass{aa}

\newcommand{\msun}{\,\mathrm{M}_\odot}

\newcommand{\au}{\,\mathrm{AU}}

\newcommand{\Myr}{\,\mathrm{Myr}}
\newcommand{\yr}{\,\mathrm{yr}}

\def\lesssim{\mathrel{\hbox{\rlap{\hbox{\lower3pt\hbox{$\sim$}}}\hbox{\raise2pt\hbox{$<$}}}}}
\def\gtrsim{\mathrel{\hbox{\rlap{\hbox{\lower3pt\hbox{$\sim$}}}\hbox{\raise2pt\hbox{$>$}}}}}
\def\gtreq{\mathrel{\hbox{\rlap{\hbox{\lower3pt\hbox{$-$}}}\hbox{\raise2pt\hbox{$>$}}}}}

\usepackage{xcolor}
\usepackage{graphicx}
\usepackage{subcaption}
\usepackage{txfonts}
\usepackage{natbib,twoopt}
\usepackage{amsmath}
\usepackage[breaklinks=true]{hyperref}%% to avoid \citeads line fills
\usepackage[T1]{fontenc}
\usepackage{cleveref}
\bibpunct{(}{)}{;}{a}{}{,}             %% natbib format for A&A and ApJ
\makeatletter
  \newcommandtwoopt{\citeads}[3][][]{\href{http://adsabs.harvard.edu/abs/#3}%
    {\def\hyper@linkstart##1##2{}%
     \let\hyper@linkend\@empty\citealp[#1][#2]{#3}}}
  \newcommandtwoopt{\citepads}[3][][]{\href{http://adsabs.harvard.edu/abs/#3}%
    {\def\hyper@linkstart##1##2{}%
     \let\hyper@linkend\@empty\citep[#1][#2]{#3}}}
  \newcommandtwoopt{\citetads}[3][][]{\href{http://adsabs.harvard.edu/abs/#3}%
    {\def\hyper@linkstart##1##2{}%
     \let\hyper@linkend\@empty\citet[#1][#2]{#3}}}
  \newcommandtwoopt{\citeyearads}[3][][]%
    {\href{http://adsabs.harvard.edu/abs/#3}
    {\def\hyper@linkstart##1##2{}%
     \let\hyper@linkend\@empty\citeyear[#1][#2]{#3}}}
\makeatother
\hypersetup{draft}

\begin{document}

   \title{The dependence of episodic accretion on eccentricity during the formation of binary stars}
   \titlerunning{Episodic accretion in binary star formation}

   %\subtitle{I. Overviewing the $\kappa$-mechanism}

   \author{Rajika L. Kuruwita
          \inst{1},
          Christoph Federrath\inst{2},
          \and
          Troels Haugb{\o}lle\inst{1}
          }

   \institute{Niels Bohr Institute,
              University of Copenhagen, {\O}ster Voldgade 5-7, DK-1350, Copenhagen K, Denmark\\
              \email{rajika.kuruwita@nbi.ku.dk}
         \and
             Research School of Astronomy and Astrophysics, Australian National University, Canberra, ACT 2611, Australia\\
             }

   \date{Received September 15, 1996; accepted March 16, 1997}

% \abstract{}{}{}{}{} 
% 5 {} token are mandatory
 
  \abstract
  % context heading (optional)
  % {} leave it empty if necessary  
   {Episodic accretion has been observed in short-period binaries, where bursts of accretion occur at periastron. The binary trigger hypothesis has also been suggested as a driver for accretion during protostellar stages.}
  % aims heading (mandatory)
   {Our goal is to investigate how the strength of episodic accretion bursts depends on eccentricity.}
  % methods heading (mandatory)
   {We investigate the binary trigger hypothesis in longer-period ($>20\yr$) binaries by carrying out three-dimensional magnetohydrodynamical (MHD) simulations of the formation of low-mass binary stars down to final separations of $\sim$10$\au$, including the effects of gas turbulence and magnetic fields. We ran two simulations with an initial turbulent gas core of one solar mass each and two different initial turbulent Mach numbers, $\mathcal{M} = \sigma_v/c_s = 0.1$ and $\mathcal{M}=0.2$, for $6500\yr$ after protostar formation.}
  % results heading (mandatory)
   {We observe bursts of accretion at periastron during the early stages when the eccentricity of the binary system is still high. We find that this correlation between bursts of accretion and passing periastron breaks down at later stages because of the gradual circularisation of the orbits. For eccentricities greater than $e=0.2$, we observe episodic accretion triggered near periastron. However, we do not find any strong correlation between the strength of episodic accretion and eccentricity. The strength of accretion is defined as the ratio of the burst accretion rate to the quiescent accretion rate.
   %From high resolution simulations w
   We determine that accretion events are likely triggered by torques between the rotation of the circumstellar disc and the approaching binary stars. We compare our results with observational data of episodic accretion in short-period binaries and find good agreement between our simulations and the observations.}
  % conclusions heading (optional), leave it empty if necessary 
   {We conclude that episodic accretion is a universal mechanism operating in eccentric young binary-star systems, independent of separation, and it should be observable in long-period binaries as well as in short-period binaries. Nevertheless, the strength depends on the torques and hence the separation at periastron.} %Since episodic accretion is a function of orbital phase and it is independent of time, it can be used to explain the different timescales of EXor type outbursts.}


   \maketitle
%
%-------------------------------------------------------------------

\section{Introduction}

A significant fraction of stars are born in binaries or multiple star systems \citep{raghavan_survey_2010, moe_mind_2017}. Binaries of separations $\lesssim10\au$ cannot form in situ during molecular core collapse because the initial hydrostatic core that collapses to form the protostar has a radius of $\sim$5$\au$ \citep{larson_numerical_1969}, and this hydrostatic core is not susceptible to fragmentation during the second protostellar collapse phase \citep{bate_collapse_1998}. Therefore, binaries with a semi-major axis $a\lesssim10\au$ likely form via the in-spiral of an initially wider binary, possibly via viscous evolution through discs \citep{gorti_orbital_1996, stahler_orbital_2010, korntreff_towards_2012}, especially circumbinary discs \citep{artymowicz_effect_1991, pringle_properties_1991}, the Kozai-Lidov mechanism \citep{kiseleva_tidal_1998}, or dynamical interactions in clustered star formation \citep{bate_formation_2002}. The ejection of a companion may enhance or initiate these processes \citep{moe_dynamical_2018}. Turbulence and magnetic fields also play a significant role in the structure and evolution of the disc \citep{seifried_accretion_2015, kuffmeier_zoom-simulations_2017, gerrard_role_2019}.

During viscous evolution of the gas disc, the angular momentum of the binary can be transferred to the gas, thus shrinking the orbit of the binary. The binary system may harden to a separation at which material in circumstellar discs is redistributed to form one circumbinary disc \citep{reipurth_fu_2004, kuruwita_role_2019}.

During this dynamical evolution, accretion events may be triggered. Triggered accretion has been observed in short-period binaries such as TWA 3A \citep[$34.8\,\mathrm{day}$,][]{tofflemire_pulsed_2017} and DQ Tau \citep[$15.8\,\mathrm{day}$,][]{tofflemire_accretion_2017}, where the accretion rate at periastron is approximately three times greater than the quiescent rate. There is little observational data on episodic accretion in long-period binaries, but the `binary trigger' hypothesis \citep{bonnell_binary_1992} has been proposed as the trigger of FU Orionis \citep[time scale $\sim10-100\yr$,][]{hartmann_fu_1996} and EXor \citep[$\sim1\yr$,][]{herbig_1993-1994_2001} type outbursts.

Understanding accretion behaviour in binary systems is necessary to comprehend the formation and evolution of discs around in binaries and hence the formation of planets in binary-star systems. The presence of a companion can truncate discs leading to faster erosion via dynamical interactions \citep[of the order of $\sim$0.3$\Myr$][]{williams_protoplanetary_2011}. However, there also exist circumbinary discs with ages greater than the typical disc lifetime of $3\Myr$ \citep{haisch_disk_2001, mamajek_initial_2009}, and this may be due to lower photo-evaporation rates in binaries \citep{alexander_dispersal_2012}. Examples of old circumbinary discs include HD 98800 B \citep[$8.5\pm1.5\Myr$][]{ducourant_tw_2014} and AK Sco \citep[$18\pm1\Myr$][]{czekala_disk-based_2015}. Overall, the influence of multiplicity on the disc lifetime has yet to be determined. Shorter circumstellar and circumbinary disc lifetimes are implied by the low disc frequency around binaries of separation $a<40\au$ \citep{cieza_primordial_2009, duchene_planet_2010, kraus_mapping_2011}; however, \citet{kuruwita_multiplicity_2018} find that overall, the lifetime of discs in binaries may not vary significantly compared to disc lifetimes around single stars.

In this paper we explore episodic accretion seen during the formation of binary stars, similar to the simulations presented in \citet{kuruwita_role_2019}. We advanced the previous simulations to $6500\yr$ after protostar formation and performed them at a higher resolution. Here, we also study how accretion behaviour evolves with eccentricity.

In \Cref{sec:method} we describe the simulation code used, how protostar formation is modelled, and our simulation setup. The results are presented and discussed in \Cref{sec:results}, where we analyse the evolution of the binary systems, study the accretion behaviour and determine the mechanism that triggers an accretion event, and compare that with observations. \Cref{sec:caveats} discusses the limitations and caveats of this study. Our conclusions are summarised in \Cref{sec:conclusion}.

% ======================================================
\section{Method}
\label{sec:method}

\subsection{The \texttt{FLASH} code}
\label{ssec:flash}

We use the adaptive mesh refinement (AMR) code \texttt{FLASH} \citep{fryxell_flash:_2000, dubey_challenges_2008} to integrate the compressible ideal MHD equations. Here we use the HLL3R Riemann solver for ideal MHD \citep{waagan_robust_2011}. The gravitational interactions of the gas are calculated using a tree-based Poisson solver \citep{wunsch_tree-based_2018}.

Our simulations use a piecewise polytropic equation of state, given by
\begin{equation}
P_\mathrm{th} = K\rho^\Gamma,
\label{eqn:eos}
\end{equation}
where $K$ is the polytropic coefficient and $\Gamma$ is the polytropic index; $K$ is given by the isothermal sound speed squared. In our simulations, the sound speed is initially set to $c_\mathrm{s} = 2 \times 10^4\,\mathrm{cm}\,\mathrm{s}^{-1}$ for a gas
temperature of $\sim11\,\mathrm{K}$ with mean molecular weight of
$2.3\,m_\mathrm{H}$, where $m_\mathrm{H}$ is the mass of a hydrogen atom. $K$ is then subsequently computed, such that $P$ is a continuous function of $\rho$. For our simulations $\Gamma$ is defined as
\begin{equation}
\Gamma=\begin{cases}
1.0  \text{ for \,\,\,\,\,\,\,\,\,\,\,\, $\rho \leq \rho_1 \equiv 2.50 \times 10^{-16}\,\mathrm{g}\,\mathrm{cm}^{-3}$},\\
1.1  \text{ for $\rho_1 < \rho \leq \rho_2 \equiv 3.84 \times 10^{-13}\,\mathrm{g}\,\mathrm{cm}^{-3}$},\\
1.4  \text{ for $\rho_2 < \rho \leq \rho_3 \equiv 3.84 \times 10^{-8}\,\mathrm{g}\,\mathrm{cm}^{-3}$},\\
1.1  \text{ for $\rho_3 < \rho \leq \rho_4 \equiv 3.84 \times 10^{-3}\,\mathrm{g}\,\mathrm{cm}^{-3}$},\\
5/3 \text{ for \,\,\,\,\,\,\,\,\,\,\,\,$\rho > \rho_4$}.
\end{cases}
\label{eqn:gamma}
\end{equation}
The values of $\Gamma$ were based on radiation-hydrodynamical simulations of molecular-core collapse by \citet{masunaga_radiation_2000}. These values approximate the gas behaviour during the initial isothermal collapse of the molecular core, adiabatic heating of the first core, the H$_2$ dissociation during the second collapse into the second core and the return to adiabatic heating. 

The formation of sink particles indicates the formation of a protostar \citep{federrath_modeling_2010, federrath_implementing_2011, federrath_modeling_2014}. A second-order leapfrog integrator is used to update the sink particle positions with a variable time step. To prevent artificial precession of the sink particles, a sub-cycling method is implemented \citep{federrath_modeling_2010}. The interactions between sink particles and the gas are computed using direct $N$-body evaluation of the forces.

\subsection{Simulation setup}
\label{ssec:simulationsetup}

The simulation methods are identical to the simulations of \citet{kuruwita_role_2019}. Here we only summarise the main elements of the method and refer the reader to \citet{kuruwita_role_2019} for the details. We simulate the formation of a binary star with an initially turbulent velocity field.

The size of the three-dimensional computational domain is $\ell_\mathrm{box}=1.2\times10^{17}\,\mathrm{cm}$ ($\sim$8000$\au$) along each side of the Cartesian domain. We use 12 levels of refinement ($L_\mathrm{ref}$) of the AMR grid, resulting in a minimum cells size of $\sim$1.95$\au$ when fully refined. At this resolution the accretion radius of the sink particles is $r_\mathrm{sink}\sim$4.9$\au$. A resolution study was conducted and is discussed in \Cref{sec:appendix}. This resolution study shows that $L_\mathrm{ref}=12$ is suitable for running long simulations and to study the accretion behaviour as a function of eccentricity with sufficient resolution in eccentricity space.

Our simulations begin with a spherical cloud of mass $1\msun$, and radius $\sim$3300$\au$ placed in the centre of the simulation domain. The cloud is initially given solid body rotation with angular momentum of $1.85\times10^{51}\,\mathrm{g}\,\mathrm{cm}^2\,\mathrm{s}^{-1}$. With this angular momentum, the product of the angular frequency and the freefall time of the cloud is $\Omega\times\,t_\mathrm{ff}=0.2$ (see \cite{banerjee_outflows_2006} and \cite{machida_high-_2008}). A initially uniform magnetic field of $100\,\mu\mathrm{G}$ is also threaded through the cloud in the \emph{z}-direction. This gives a mass-to-flux ratio of $(M/\Phi)/(M/\Phi)_{\mathrm{crit}}=5.2$ where the critical mass-to-flux ratio is $487\,\mathrm{g}\,\mathrm{cm}^{-2}\,\mathrm{G}^{-1}$ as defined in \citet{mouschovias_note_1976}. 
The cloud is initially given a uniform density of $\rho_0=3.82\times10^{-18}\,\mathrm{g}\,\mathrm{cm}^{-3}$, and a density perturbation is imposed on the cloud. This is to seed the formation of a binary-star system. While observations find a bi-modal distribution in the separation of pre-main sequence binaries, likely due to formation via core and disc fragmentation \citep{tobin_vla/alma_2018}, our simulations focus on the core fragmentation pathway. These observations find that the companion frequency is higher for wider binaries, suggesting that core fragmentation is more likely. This is concurrent with previous theoretical work of multiple star formation from a single molecular core, which also suggest that core fragmentation is a more likely pathway of multiple star formation rather than disc fragmentation \citep{offner_formation_2010}, because radiation feedback increases the Jeans length within discs, which tends to suppress disc fragmentation. The density perturbation in our simulations is described by
\begin{equation}
\rho = \rho_0 (1 + \alpha_\mathrm{p}\mathrm{cos}\varphi),
\label{eqn:densityperturbation}
\end{equation}
where $\varphi$ is the angle about the $z$-axis and $\alpha_\mathrm{p}$ is the amplitude of the perturbation. For our simulations $\alpha_\mathrm{p}=0.5$. This perturbation is a standard method of seeding binary star formation within simulations of molecular cores \citep{boss_fragmentation_1979, bate_resolution_1997, kuruwita_binary_2017}.

In order to prevent the cloud from expanding rapidly, the spherical cloud is in pressure equilibrium with the surrounding material. This is achieved by giving the surrounding material a gas density of $\rho_0/100$ with an internal energy such that the cloud and surrounding material is in pressure equilibrium. The inflow and outflow boundary conditions are used at the edge of our computational domain.

An initial turbulent velocity field is imposed on top of the solid body rotation. We run two simulations with turbulence of Mach number $\mathcal{M}=0.1$ and $0.2$, which are referred to as \emph{T1} and \emph{T2} hereafter. For details on the implementation of turbulence we refer the reader to \citet{federrath_comparing_2010} and \citet{kuruwita_role_2019}.

\begin{figure*}
	%\centerline{\includegraphics[width=1.0\linewidth]{Images/Side_on_volume_weighted.pdf}}
	\vspace{-0.3cm}
	\centerline{\includegraphics[width=1.0\linewidth]{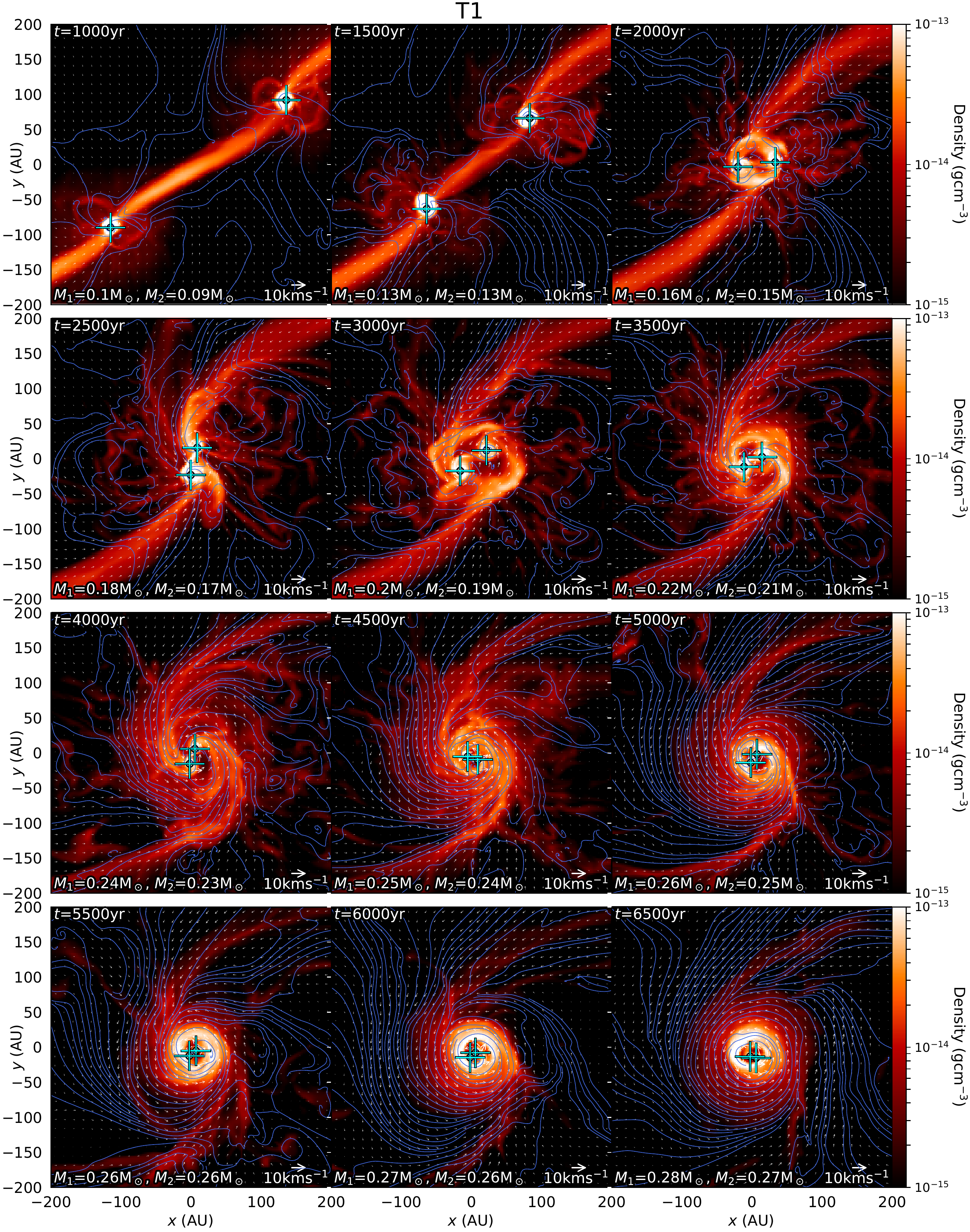}}
	\vspace{-0.3cm}
	\caption{$300\au$ thick volume-weighted slices through the gas density orientated along the $z=0$ plane (perpendicular to the rotation axis of the core) for \emph{T1}. Each panel progresses at $500\yr$ intervals since $1000\yr$ after the first protostar formation. The thin lines show the magnetic field, and the arrows indicate the velocity field. Crosses show the position of the sink particles. The mass accreted by the sink particles in the simulations is indicated on the bottom left of each panel. The circles around the centre of the crosses indicate the sink particle accretion radius.}
	\label{fig:slices_T1}
\end{figure*}

\begin{figure*}
	%\centerline{\includegraphics[width=1.0\linewidth]{Images/Side_on_volume_weighted.pdf}}
	\vspace{-0.4cm}
	\centerline{\includegraphics[width=1.0\linewidth]{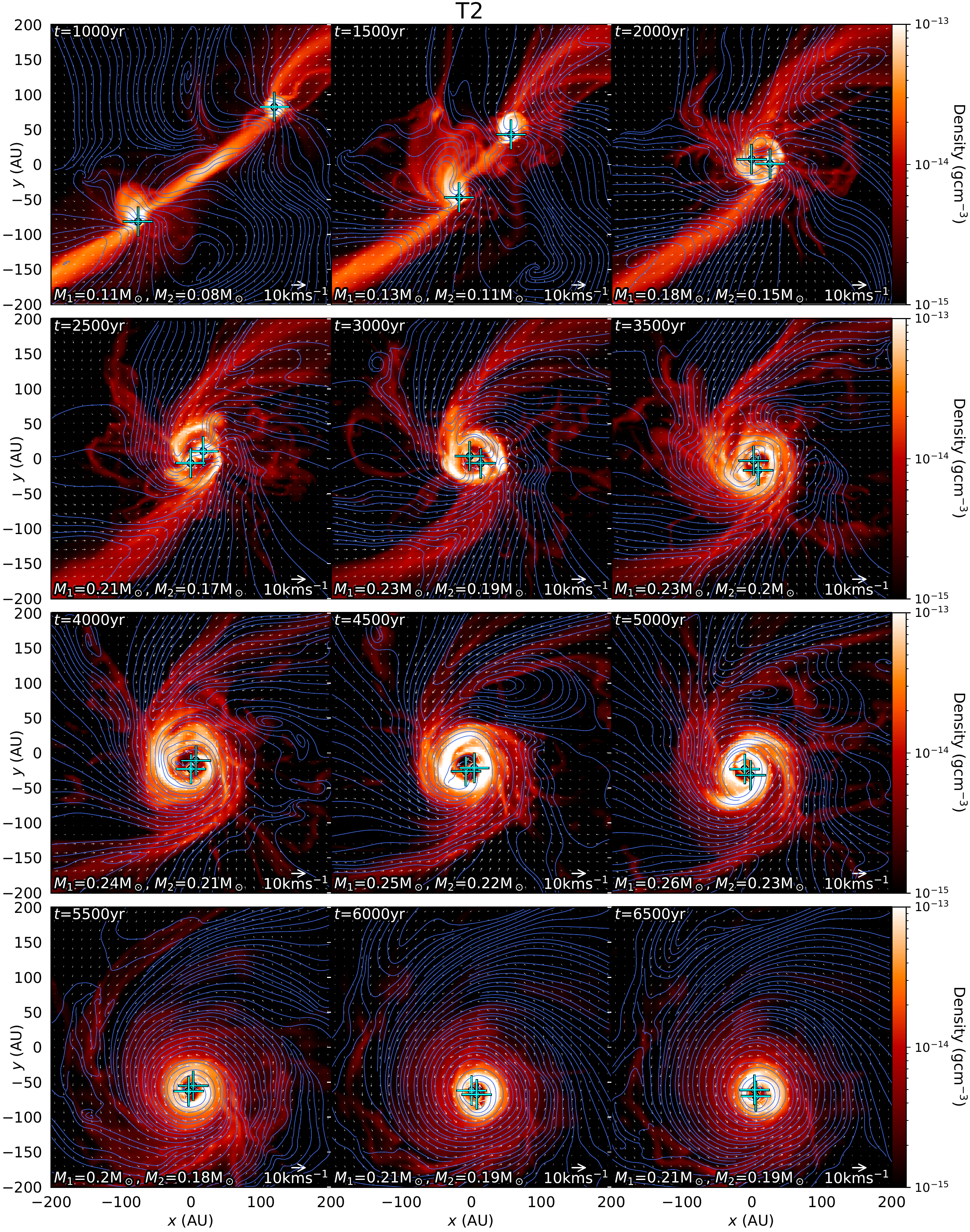}}
	% \vspace{-0.4cm}
	\caption{Same as \Cref{fig:slices_T1}, but for \emph{T2}.}
	\label{fig:slices_T2}
\end{figure*}

\begin{figure}
	%\centerline{\includegraphics[width=1.0\linewidth, trim={4mm, 0.0mm, 10.0mm, 0.0mm}]{Images/outflow_quantities.pdf}}
	\centerline{\includegraphics[width=1.0\linewidth]{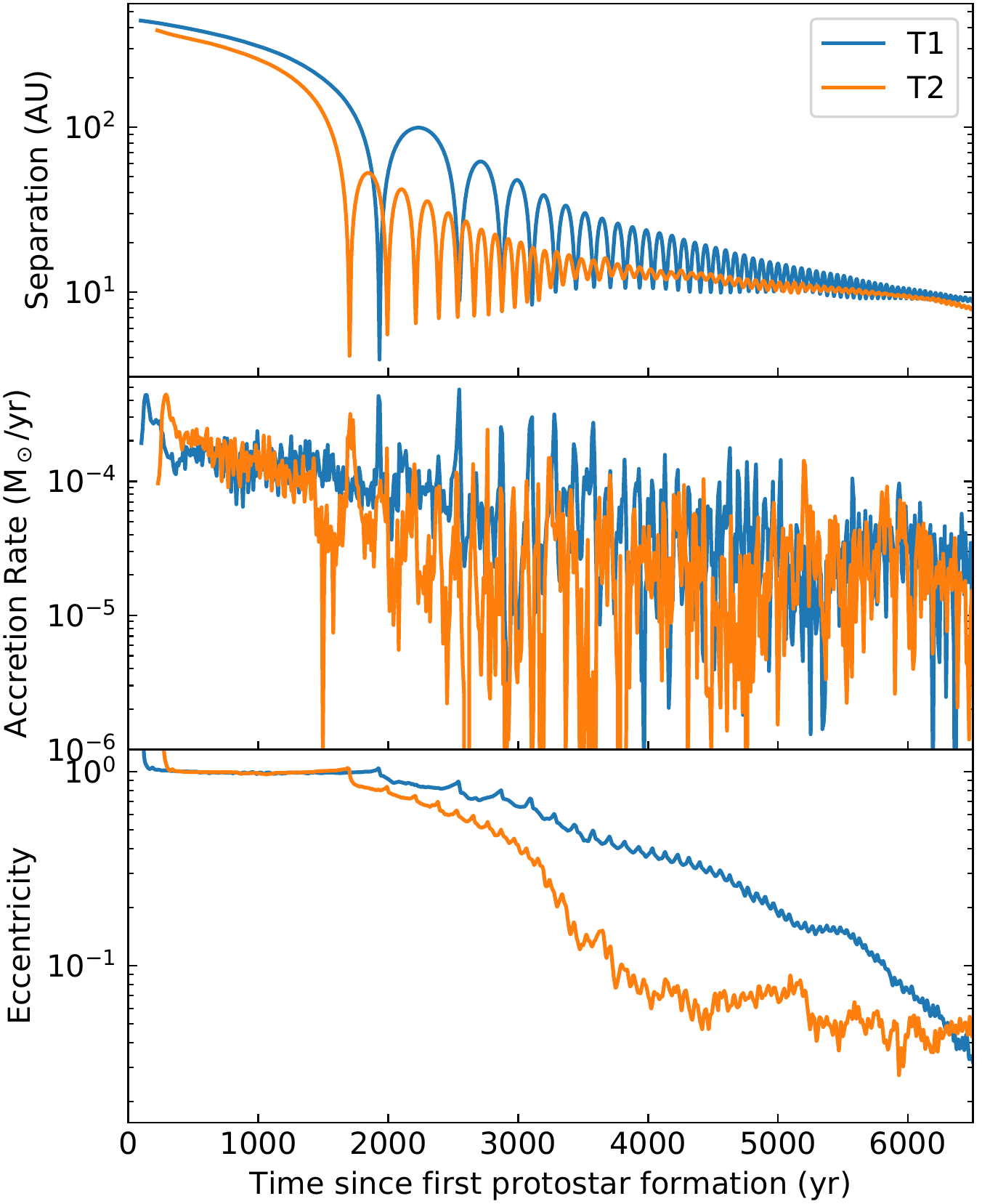}}
	\caption{Evolution of the binary separation (top), accretion rate (middle) and eccentricity (bottom) since protostar formation of the first sink particle for the \emph{T1} and \emph{T2} cases in blue and orange, respective. At early times, high accretion rates are clearly correlated with periastrons, in both \emph{T1} and \emph{T2}.}
	\label{fig:SystemEvolution}
\end{figure}

% =====================================
\section{Results and discussion}
\label{sec:results}

The simulations were run for $\sim$6500$\yr$ after the formation of the first protostar. Top-down density slices showing the disc and binary system face-on of \emph{T1} and \emph{T2} are shown in Figures~\ref{fig:slices_T1} and \ref{fig:slices_T2}, respectively.

\subsection{Time evolution of the binary-star system}
\label{ssec:evolution}

As the simulations progress the spherical cloud collapses and sink particles are created in collapsing regions along a filament that forms as a result of the initial density perturbation. Sink particles form at separations between $400$ and $500\au$ and fall towards the centre of the initial dense core as shown in Figures~\ref{fig:slices_T1} and~\ref{fig:slices_T2}. These binary systems are evolved for $6500$ years after the formation of the first sink particle, which ensures that the binary is able to complete many orbits to form an established binary system of semi-major axis \mbox{$9$--$10\au$} in both simulations.

In \Cref{fig:SystemEvolution} we show the separation, total accretion rate (the sum of the accretion rate of the primary and secondary components), and eccentricity evolution for \emph{T1} and \emph{T2}. The binary systems begin to establish their orbits approximately $\sim$2000$\yr$ after the formation of the first sink particles in both cases. \emph{T1} in-spirals from a $\sim$600$\yr$ orbit to a $\sim$40$\yr$ orbit, while \emph{T2} in-spirals from a $\sim$300$\yr$ orbit to a $\sim$20$\yr$ orbit. Based on the resolution study presented in \Cref{sec:appendix}, we find that orbital shrinkage occurs over a shorter period of time with increasing resolution, with the $L_\mathrm{ref} = 13$ test run being nearly converged. However, since the higher resolution simulations are much more costly in terms of computational resources, we cannot run them for very long, and we therefore focus primarily on the $L_\mathrm{ref} = 12$ for most of the following analyses, unless stated otherwise.

The separation and accretion rate are determined directly from the sink particle data. The eccentricity is calculated using
\begin{equation}
e = \sqrt{1 + \frac{2\epsilon h^2}{(GM_{tot})^2}},
\label{eqn:eccentricity}
\end{equation}
where $\epsilon$ and $h$ are the specific orbital energy (sum of kinetic and gravitational potential) and angular momentum of the system, $G$ is the gravitational constant and $M_{tot}$ is the total mass of the binary. $\epsilon$ and $h$ are calculated using
\begin{equation}
\epsilon = \frac{E_{pot} + E_{kin}}{\mu}\quad\mathrm{and}\quad h = \frac{L}{\mu},
\label{eqn:specificEnAndAng}
\end{equation}
respectively, where $E_{pot}$ and $E_{kin}$ are the orbital potential and kinetic energy, respectively, $L$ is the total angular momentum of the binary, and $\mu$ is the reduced mass, given by
\begin{equation}
\mu = \frac{M_1M_2}{M_1 + M_2},
\label{eqn:reducedMass}
\end{equation}
where $M_1$ and $M_2$ are the mass of the primary and secondary components, respectively.

As the binary-star systems evolve, they accrete mass. During the early in-spiral of the binaries, we see clear spikes in accretion correlated with the periastron passage of the binaries in \Cref{fig:SystemEvolution}. These accretion events are less prominent at later stages of the binary evolution when the binaries have evolved to a lower eccentricity, and there is less circumstellar material. This episodic accretion supports the binary-trigger hypothesis for accretion outbursts \citep{bonnell_binary_1992, green_testing_2016}. Previous work on binary-star accretion from circumbinary discs find some dependence on the eccentricity of the inner binary \citep{gunther_circumbinary_2002, miranda_viscous_2017, munoz_hydrodynamics_2019}, with binaries of higher eccentricity showing this episodic accretion, while circular binaries did not show episodic accretion. These previous simulations, however, artificially drive the eccentricity of the inner binary and begin with a Class II disc. The systems produced from our simulations evolve naturally from the collapse of a molecular cloud core. Since our binary systems evolve through a range of eccentricities, we investigate this episodic accretion as a function of the eccentricity evolution of these binaries.

\subsection{Accretion events and dependence on eccentricity}
\label{ssec:episodic_accretion}

Now we investigate the accretion behaviour as a function of orbital phase and eccentricity. In order to phase-fold the accretion we identify the times of periastron and apastron. The time of periastron and apastron are defined as orbital phase, $\phi=0$ and $0.5$, respectively. The sink particle data is then divided into ten time bins between each periastron and apastron, which results in a total of 20 time bins or phase-space bins between consecutive periastrons. In each bin, the average accretion rate $\langle\dot{M}\rangle$ is calculated via
\begin{equation}
\langle\dot{M}\rangle=\frac{\int_{t_{bin}}^{t_{bin+1}}\dot{M}(t)dt}{\int_{t_{bin}}^{t_{bin+1}}dt},
\label{eqn:averagedaccretion}
\end{equation}
where $\dot{M}$ is the accretion rate at time $t$, $t_{bin}$ and $t_{bin+1}$ are the bounds of the bin, and $dt$ is the simulation time step.

\begin{figure*}                                                       
	%\centerline{\includegraphics[width=1.0\linewidth, trim={4mm, 0.0mm, 10.0mm, 0.0mm}]{Images/outflow_quantities.pdf}}
	\centerline{\includegraphics[width=1.0\linewidth]{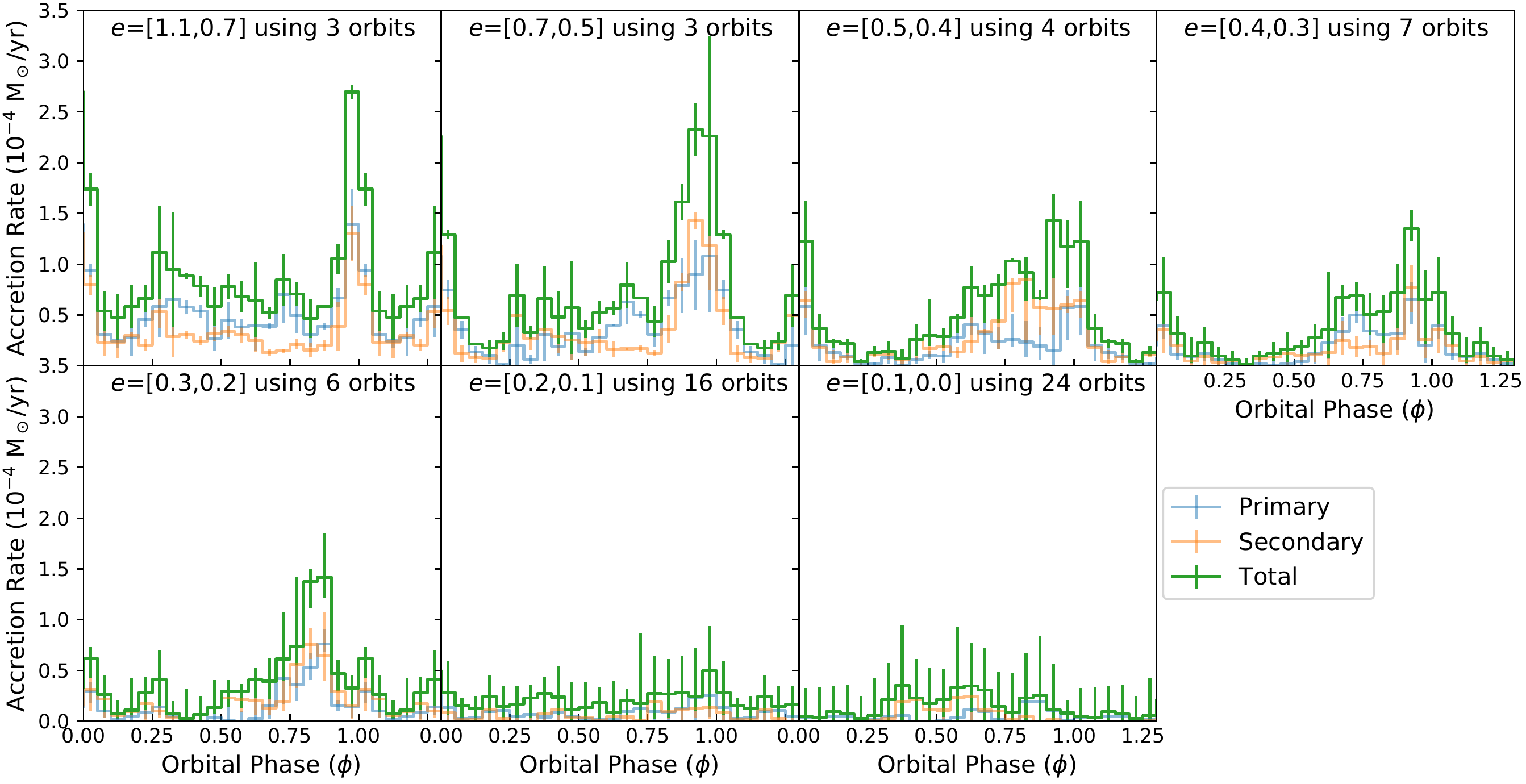}}
	\caption{Shows phase-folded accretion within the defined eccentricity bins (see \Cref{tab:simulation_summary}) for \emph{T1}. The transparent blue and orange lines are the median accretion rate for the primary and secondary components of the binary, respectively. The solid green line is the total accretion rate. The error for each bin is taken to be the 16th and 84th percentile.}
	\label{fig:componentAccretion1}
\end{figure*}

After the phase-folding is completed the median accretion rate in each phase-space bin is calculated and the orbit-to-orbit variation for each bin is taken to be the 16th and 84th percentile. Since the eccentricity of the binary system is also evolving over the course of the simulations, it is not appropriate to phase-fold the accretion over the entire duration of the simulation. In order to study the dependence of the intensity of episodic accretion on eccentricity, we define eccentricity bins to phase-fold over. For \emph{T1}, these eccentricity bins were selected to be $e=\;$[1.1, 0.7, 0.5, 0.4, 0.3, 0.2, 0.1, 0.0]. As the eccentricity in \emph{T2} reduces at a faster rate than in \emph{T1}, these same bins were not appropriate for \emph{T2}, because in some bins only two orbits would be folded. Thus, for \emph{T2}, we adjusted the bins to be $e=\;$[1.1, 0.7, 0.5, 0.3, 0.1, 0.0]. This is summarised in \Cref{tab:simulation_summary}. We identified which periastrons fell into each eccentricity bin and then phase-folded the orbits within each eccentricity bin.

\begin{table}
	\caption{Summary of the simulation analysis. The left and middle columns give the simulation name and turbulent Mach number. The right column lists the eccentricity bins used for the episodic accretion analysis.}
	\label{tab:simulation_summary}
	\centering
	\begin{tabular}{lccc}
		\hline
		Simulation & $\mathcal{M}$ & $e$ bins\\
		\hline
		\emph{T1} & $0.1$ & [1.1, 0.7, 0.5, 0.4, 0.3, 0.2, 0.1, 0.0] \\
		\emph{T2} & $0.2$ & [1.1, 0.7, 0.5, 0.3, 0.1, 0.0] \\ 
		\hline
	\end{tabular}
\end{table}

\begin{figure*}
	%\centerline{\includegraphics[width=1.0\linewidth, trim={4mm, 0.0mm, 10.0mm, 0.0mm}]{Images/outflow_quantities.pdf}}
	\centerline{\includegraphics[width=1.0\linewidth]{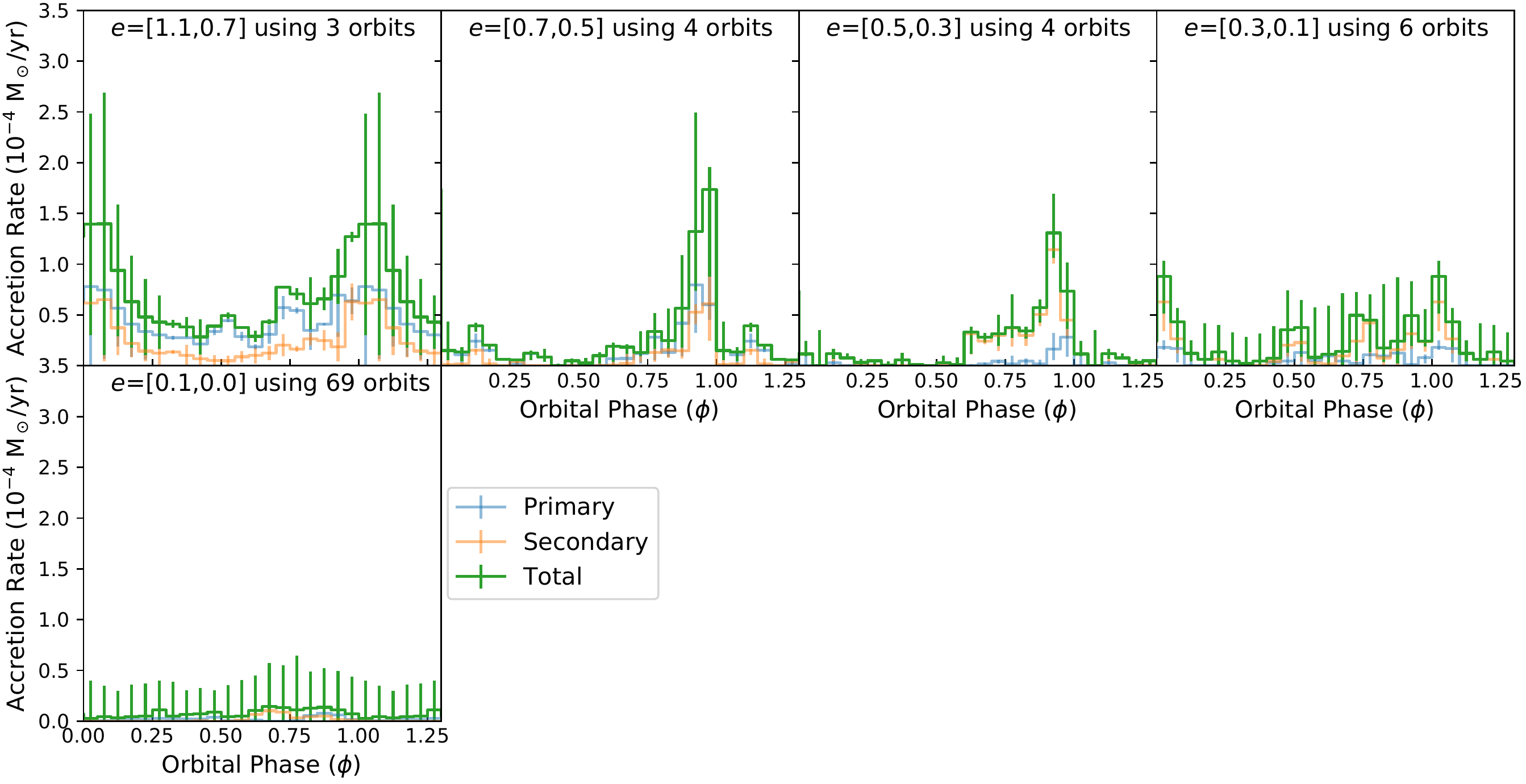}}
	\caption{Same as \Cref{fig:componentAccretion1}, but for \emph{T2}.}
	\label{fig:componentAccretion2}
\end{figure*}

We show the accretion rate as a function of orbital phase for the various eccentricity bins in Figures~\ref{fig:componentAccretion1} and~\ref{fig:componentAccretion2}, for \emph{T1} and \emph{T2}, respectively. We also annotate how many orbits were in each eccentricity bin. The solid green line shows the total accretion rate, while the transparent blue and orange lines show the phase-folded accretion rate for the primary and secondary components of the binary, respectively. Based on the resolution study presented in \Cref{sec:appendix}, we demonstrate that the maximum accretion rate is converged for eccentricities of $e\lesssim0.6$, but increases with resolution for $e\gtrsim0.6$. Therefore, the maximum accretion rates represented here for high eccentricities should be taken as a lower limit.

From the phase-folded accretion we see very prominent accretion events happening near periastron, when the phase is between $\phi=$ 0.8 and 1.1, above an eccentricity of around \mbox{$e=0.2$--$0.3$}. For \emph{T1} we see a clear trend in the absolute accretion rate at periastron being larger for higher eccentricities. In \emph{T1} we do not see a clear preference for the primary or secondary being the stronger accretor over all eccentricities. We do see that for $e=[1.1, 0.7]$, the primary component accretes at a higher rate during the quiescent phases. The secondary also appears to have stronger episodic accretion for $e=[1.1,0.5]$. Simulations of accretion in circular binaries have found that the secondary components are the main accretors from protostellar envelopes and discs \citep{bate_accretion_1997, bate_predicting_2000}, as well as from circumbinary discs \citep{young_binary_2015, young_evolution_2015, munoz_hydrodynamics_2019}. The binaries in these cases have lower mass-ratios ($q<0.9$) than the binaries in our simulations ($q>0.9$), and this may play a role in which component is the primary accretor.

In \emph{T2} we see that the primary component accretes at a higher rate during the quiescent phases, and the secondary appears to display stronger episodic accretion. For lower eccentricities ($e<0.5$), the secondary is the stronger accretor. The mass ratios ($q=M_{primary}/M_{secondary}$) of the binaries formed are high ($q>0.9$) and this behaviour may differ with lower mass ratios, resulting in stronger differences in which star is the stronger accretor. \citet{munoz_hydrodynamics_2019} find for simulations of equal-mass binaries in eccentric orbits ($e=0.6$) that the primary and secondary alternate in which one receives most of the mass over precessional timescales.

In order to quantify the strength of the accretion events at periastron we calculate the ratio of the accretion during the burst to the quiescent accretion rate. We take the accretion rate at the burst ($\dot{M}_b$) to be the average accretion rate between phases $0.8<\phi<1.1$. The quiescent accretion rate ($\dot{M}_q$) is taken to be the average accretion rate between phases $0.2<\phi<0.75$. We use the following definition to quantify the strength of accretion, denoted $\beta$,
\begin{equation} 
\label{eqn:beta}
\beta = \dot{M}_{burst} / \dot{M}_{quiet}.
\end{equation}
The variation in $\beta$ is calculated via error propagation of the uncertainties in each phase-folded bin. For $\beta\sim1$, there is no episodic accretion, while $\beta\gg1$ indicates strong episodic accretion. The variation in eccentricity it taken to be the 16th and 84th percentile of the eccentricities within each eccentricity bin.

\begin{figure}
	%\centerline{\includegraphics[width=1.0\linewidth, trim={4mm, 0.0mm, 10.0mm, 0.0mm}]{Images/outflow_quantities.pdf}}
	\centerline{\includegraphics[width=1.0\linewidth]{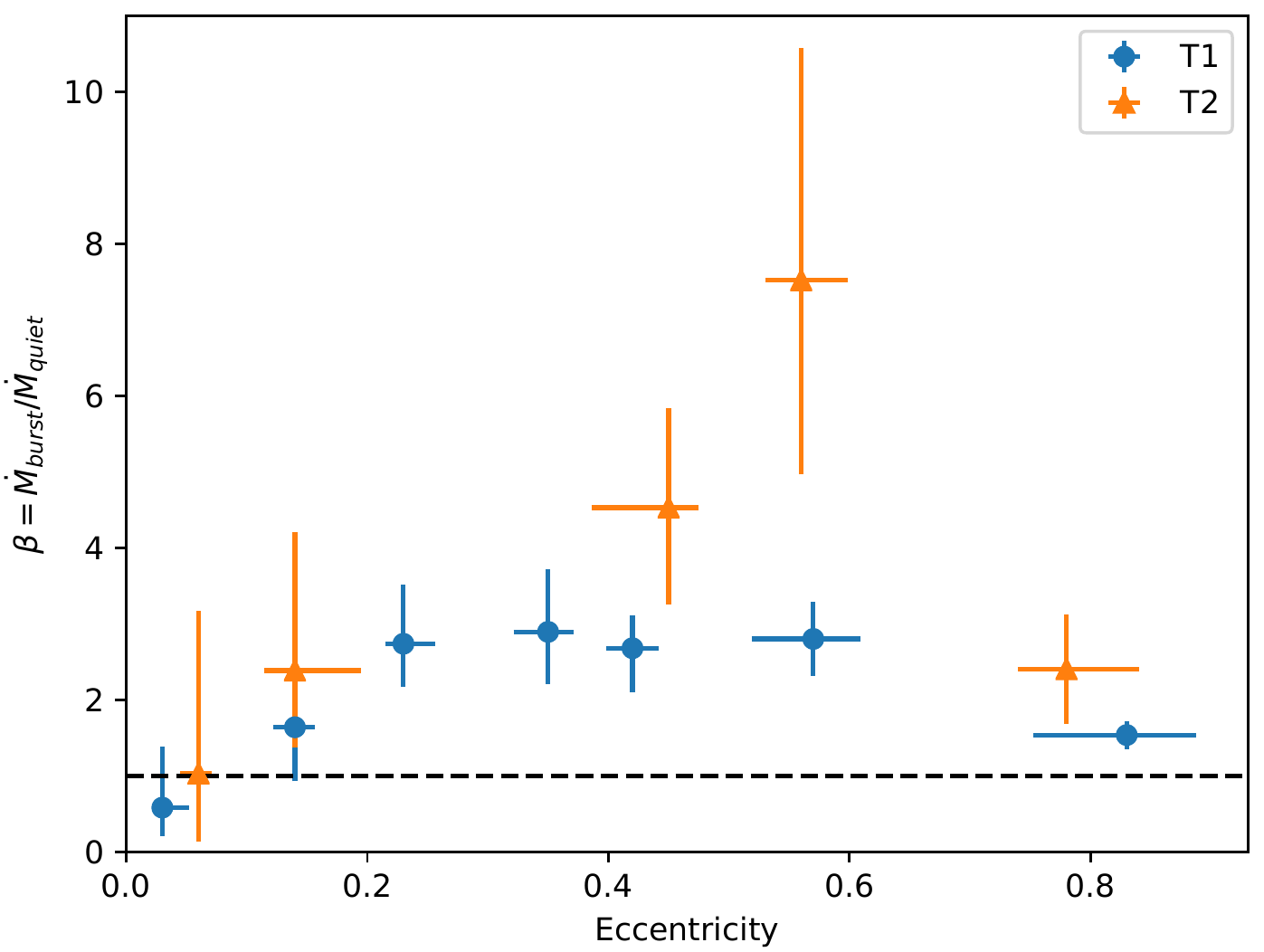}}
	\caption{Ratio of burst accretion rates to quiescent accretion rates, $\beta$, defined in \Cref{eqn:beta}, as a function of eccentricity for \emph{T1} (blue circles) and \emph{T2} (orange triangles). Episodic accretion is strongest for \mbox{$e\sim0.2$--$0.6$}.}
	\label{fig:beta}
\end{figure}

\begin{figure*}
\begin{subfigure}{\textwidth}
  \centering
  \includegraphics[width=0.95\textwidth]{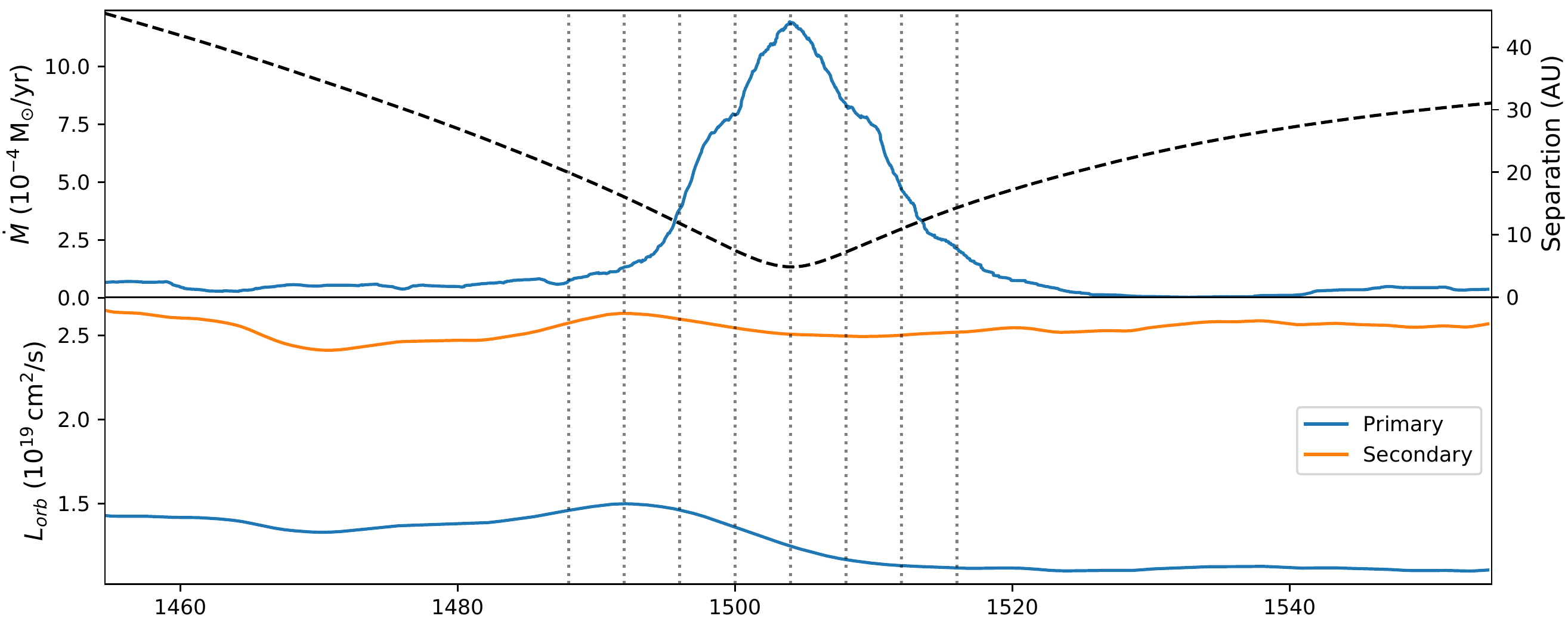}
  \label{fig:sfig1}
  %\vspace{-0.5cm}
\end{subfigure}
\begin{subfigure}{\textwidth}
  %\hspace{-0.5cm}
  \centering
  \includegraphics[width=\textwidth]{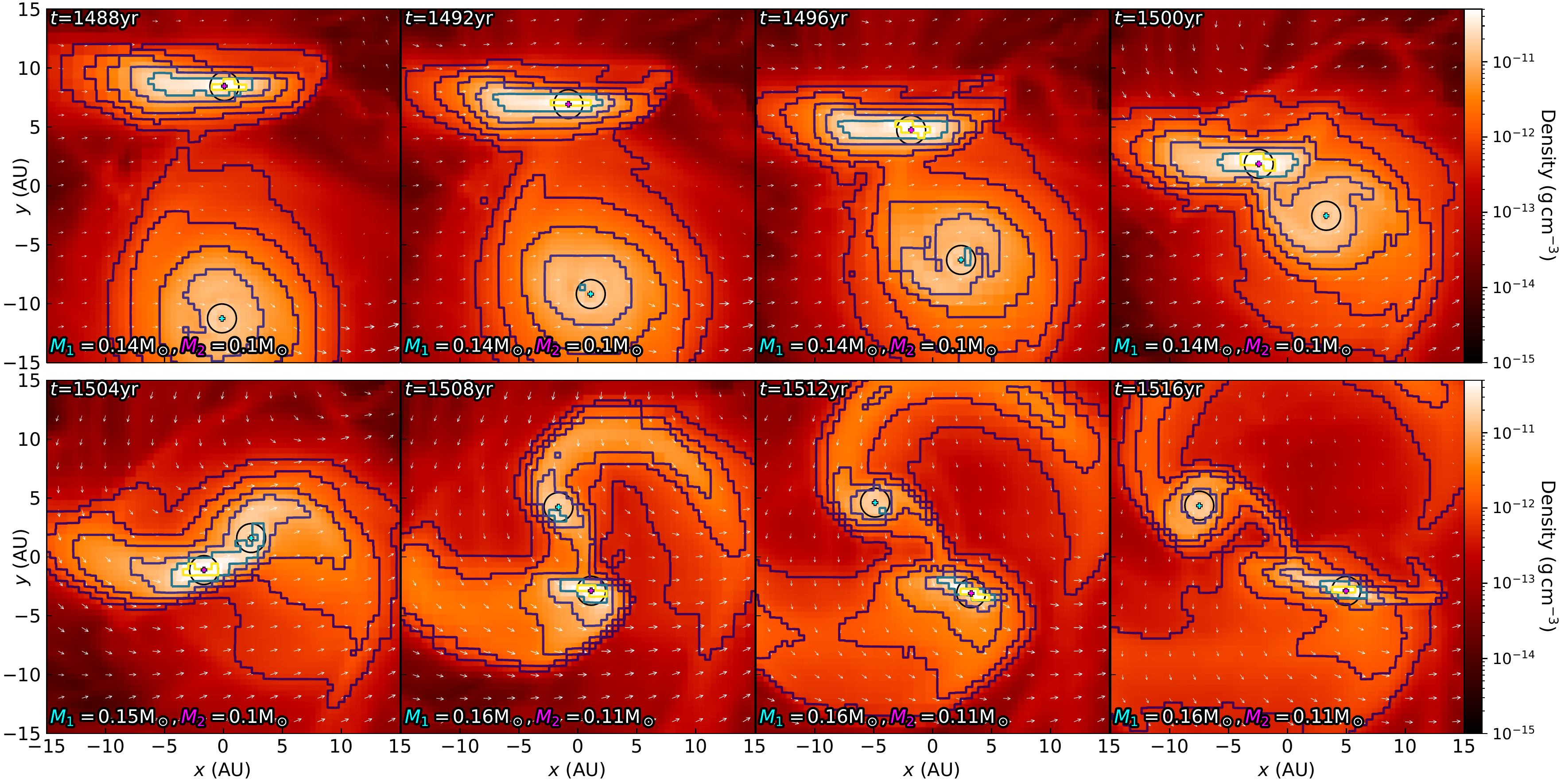}
  \label{fig:sfig2}
  \vspace{-0.4cm}
\end{subfigure}
\caption{\emph{Top}: accretion and specific orbital angular momentum profiles of the accretion burst at the first periastron passage. The profile is smoothed by taking a moving average over a window of $10\yr$. The dashed line in the top panel shows the separation of the binary. In the bottom panel the momentum is plotted for the primary and secondary star. The vertical dotted lines indicate the times when density slices were taken, which are shown in the bottom figure. \emph{Bottom}: zoomed-in slices of the high-resolution simulation of the \emph{T2} scenario with $L_\mathrm{ref}=14$. The slices are produced with the same methods as Figures~\ref{fig:slices_T1} and~\ref{fig:slices_T2} but with a projection thickness of $30\au$. The black circles show the accretion radius of the stars. The solid lines are density contours spaced evenly in log-space at density $= [0.5, 1.3, 3.2, 7.9, 20, 50]\times10^{-12}\,\mathrm{g}\,\mathrm{cm}^{-3}$.}
\label{fig:zoom_in}
\end{figure*}

In \Cref{fig:beta} we show the resulting $\beta$-values for our simulations. We find $\beta\sim1$ for eccentricities $e<0.2$, i.e. consistent with no episodic accretion. For \emph{T1}, we see that the strength of episodic accretion peaks at moderate eccentricities, $0.2<e<0.6$. A similar trend is seen for \emph{T2} with higher eccentricities displaying stronger episodic accretion up to the highest eccentricities. While \emph{T1} peaks at $\beta\sim3$, \emph{T2} produces higher $\beta$ for higher eccentricity. This may suggest some dependence of the strength of episodic accretion on the level of turbulence, with stronger turbulence producing stronger accretion. Based on the resolution study presented in \Cref{sec:appendix}, the values of $\beta$ have not completely converged at our standard resolution of $L_\mathrm{ref}=12$, generally observing higher $\beta$ for higher resolution, therefore, the values plotted in \Cref{fig:beta} should be taken as lower limits.

Interestingly, we find that for the highest eccentricities, the strength of the episodic accretion is not stronger than at moderate eccentricities, for both simulations, even if the absolute accretion rate is higher. This may be due to a combination of the quiescent accretion rate being higher at these eccentricities and accretion bursts begin shut down quickly from the circumstellar material being violently disrupted. The separation at periastron for the highest eccentricities ($a_{peri}$) is less than the radius of the circumstellar discs ($r_{disc}$). From Figures~\ref{fig:slices_T1} and \ref{fig:slices_T2} we can see that the radius of the circumstellar discs is of the order of $10\au$, while from \Cref{fig:SystemEvolution} we find that the first periastrons have separations $<10\au$ (i.e. $a_{peri}<r_{disc}$). This can lead to severe disruption of the discs, hindering efficient accretion of disc material at periastron, and promoting mass transfer between the two stellar systems.

Based on the derived $\beta$ values, we approximate that binaries with eccentricities $e\gtrsim0.2$ should display episodic accretion. However, the reason why episodic accretion is not seen at lower eccentricities in our simulations may be because at these later stages in the binary-star evolution, the circumstellar discs are significantly depleted. The reduced episodic accretion may also be partially due to limitations on the numerical resolution of the simulations, such that circumstellar disc diameters can only be resolved over approximately five cells. However, we note that the resolution study also shows no episodic accretion at low eccentricity with increasing resolution. The lack of episodic accretion at lower eccentricities is expected because of the lack of periodic variation in the gravitation potential, and hence lack of periodic forcing. Previous simulations of accretion in binaries also find that low-eccentricity binaries do not show episodic accretion \citep{munoz_pulsed_2016}.

\subsection{Mechanism driving the accretion event}

In the previous subsection we have established that accretion events are triggered near periastron for our simulated binaries. We now look in detail at the interactions to determine what physical mechanism is triggering the accretion burst.

As part of our resolution study (\Cref{sec:appendix}), we run a simulation with level of refinement of $L_\mathrm{ref}=14$ for approximately 6~orbits (see \Cref{fig:system_res}). In \Cref{fig:zoom_in} we present zoomed-in slices of the \emph{T2} simulation at various points along the accretion burst at the first periastron. In the top panel of \Cref{fig:zoom_in}, we show the zoomed-in accretion and specific orbital angular momentum evolution of this event, with the dotted line annotating the times of the slices in the panels below.

From the accretion profile and as shown in the phase-folded accretion, the accretion event begins before periastron. It is the approach of the companion that removes angular momentum from the gas in the outer disc due to the different angular velocities. This angular momentum is transferred from the gas to the orbit. We see this in the evolution of the orbital angular momentum. Prior to the accretion event, at times $t=1488$ and $1492\yr$, we see that the orbital angular momentum of the binary components increases due to the exchange with the gas. This leads to an asymmetry in the angular momentum distribution in the disc, exciting a spiral density wave. This is observed in the density contours of the slices at $t=1496$ and $1500\yr$. This excitation of spirals by a nearby companion is believed to be the cause of the observed spiral arms in systems such as HD100453 \citep{rosotti_spiral_2019}. 

It is also at times $t=1496$ and $1500\yr$ that the accretion event is reaching its peak. At periastron the circumstellar material is violently disrupted, which slows down the accretion rate. The orbital angular momentum of the binary has also been decreasing over the course of the accretion event, because it is being imparted onto the gas and ejected in spiral arms. This hardens the binary orbit while pushing gas to higher orbits, which will eventually build the circumbinary discs observed in Figures~\ref{fig:slices_T1} and~\ref{fig:slices_T2}.

\subsection{Comparison with observations}
\label{ssec:observations}

Observations of episodic accretion in binaries have typically focused on short-period binaries ($P<40\,\mathrm{day}$), because this allows for observations over multiple orbits, and hence, allows us to better understand accretion as a function of orbital phase. However, our simulations show that episodic accretion can also occur in longer-period binaries ($P>20\yr$). In this section we compare the shape of our accretion curves from long-period binaries with observed episodic accretion from the short-period binaries TWA~3A \citep[$P=34.8\,\mathrm{day}$,][]{tofflemire_pulsed_2017} and DQ~Tau \citep[$P=15.8\,\mathrm{day}$,][]{tofflemire_accretion_2017}. These binaries have eccentricities of $e=0.6280$ for TWA~3A \citep{kellogg_twa_2017}, and $e=0.568$ for DQ~Tau \citep{czekala_disk-based_2016}. They are both class~II objects, which have accretion rates significantly lower than those produced in the simulations, which are at the class~0/I stage. Therefore, to enable a meaningful comparison we compute the normalised accretion rate to study the shape of the curves. We normalise the accretion curves by dividing the accretion rate in each bin by the accretion rate averaged over all the bins. This is found by integrating the accretion rate over one orbital phase. 

\begin{figure}
	%\centerline{\includegraphics[width=1.0\linewidth, trim={4mm, 0.0mm, 10.0mm, 0.0mm}]{Images/outflow_quantities.pdf}}
	\centerline{\includegraphics[width=1.0\linewidth]{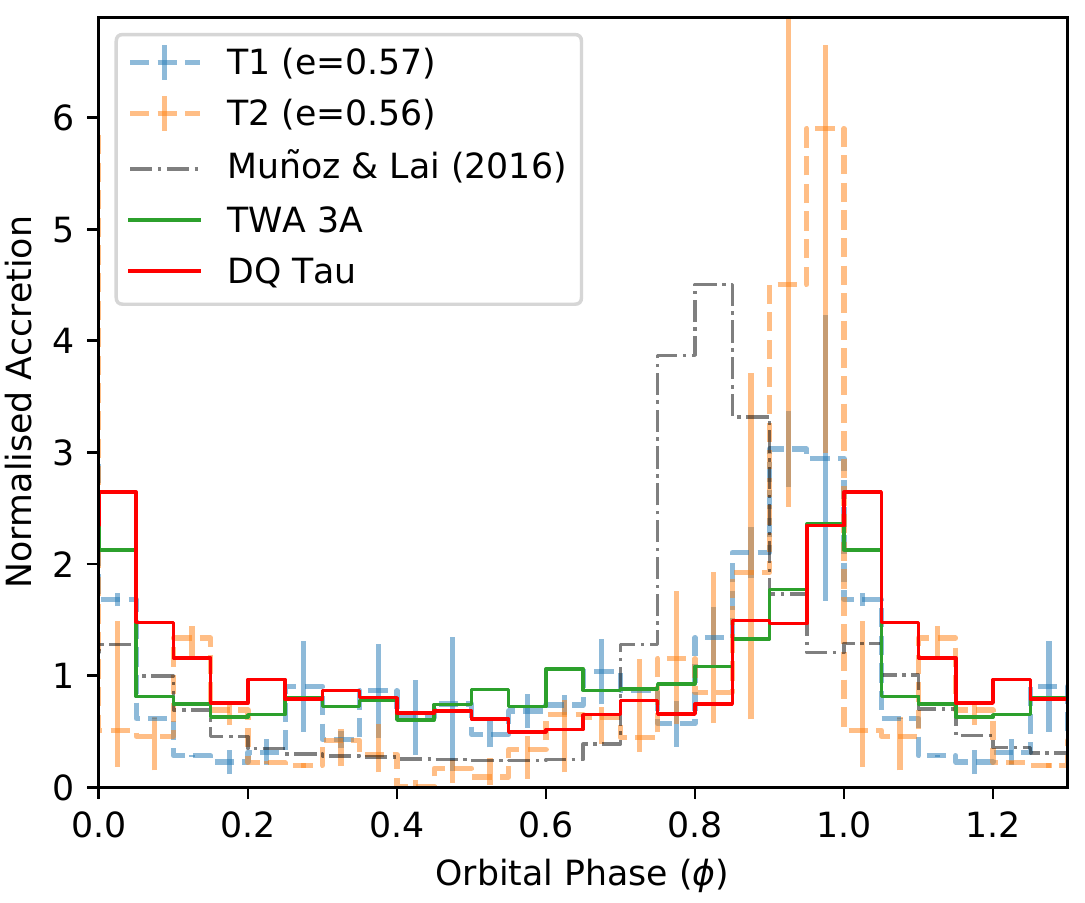}}
	\caption{Normalised observed accretion rate for TWA~3A (solid green) and DQ~Tau (solid red), against the simulated phase-folded accretion curves from \emph{T1} (dashed blue) and \emph{T2} (dashed orange), which have eccentricities similar to the observed systems. We also show the results of the simulations from \citet{munoz_pulsed_2016} for a binary of eccentricity $e=0.5$ (black dash-dotted line).}
	\label{fig:obs_comp}
\end{figure}

In \Cref{fig:obs_comp} we show the observed accretion curve against the simulated curves for the bins containing the observed eccentricity of the systems (i.e. the eccentricity bin $e=[0.7, 0.5]$). The resolution study in \Cref{sec:appendix} shows that the maximum accretion rate has mostly converged for these eccentricities, however, the quiescent accretion rate still drops slightly with increasing resolution. We find good agreement between the quiescent accretion rate of \emph{T1} and the observations, while \emph{T2} appears to produce quiescent accretion rates about 50\% lower than those observed. \emph{T1} also reproduces the observed burst accretion rate, while \emph{T2} produce a factor $\sim 2$ higher burst accretion rates, respectively, than those observed. This comparison may suggest that these class~II systems have lower turbulence, resulting in weaker episodic accretion.

We also plot the results of \citet{munoz_pulsed_2016} in \Cref{fig:obs_comp}, who ran two-dimensional, non-self-gravitating, hydrodynamic simulations of binary accretion using the \texttt{AREPO} code \citep{springel_e_2010} for a binary of eccentricity $e=0.5$. The most prominent difference between \citet{munoz_pulsed_2016} on one hand, and the observations and our simulations on the other, is that the simulation in \citet{munoz_pulsed_2016} produces the peak accretion rate significantly earlier in terms of orbital phase, namely at around $\phi=0.8$. The simulations by \citet{munoz_pulsed_2016} also overestimate the peak accretion rate by a factor $\sim$1.5 compared to the observations, similar to the \emph{T2} simulation presented here. However, the better agreement of the simulations presented here with the observations, in terms of orbital phase, may be related to the inclusion of magnetic fields, which are absent in the simulations by \citet{munoz_pulsed_2016}. Magnetic fields increase gas viscosity leading to shorter viscous timescales between disc perturbation and accretion. However, from the presented suite of simulations, it is still unclear whether magnetic fields would explain the discrepancies between the peaks of accretion.

%Given that neither \citet{munoz_pulsed_2016} nor our simulations accurately reproduce the observed peak accretion behaviour indicates that neither the inclusion of magnetic fields nor turbulence played a role in producing more realistic burst accretion rates. However, the better agreement of the simulations presented here with the observations, in terms of orbital phase, may be related to the inclusion of turbulence and magnetic fields, which are absent in the simulations by \citet{munoz_pulsed_2016}. Ultimately, a systematic comparison based on the same numerical setup evolved with different codes is needed to draw firm conclusions.

There are some differences between the observed binaries and the simulated systems. Our simulations study the evolution of class~0/I binaries with massive discs, while the observations study class~II binaries. Despite the differing evolutionary stages, there is overall good agreement between the shape of the simulated accretion curves, considering that the simulated and observed binaries have vastly different orbital periods. This suggests that the accretion behaviour at periastron for eccentric binaries is largely independent of the orbital separation and signs of episodic accretion may be detected in long-period binaries. Despite the relatively good agreement between the simulations and the observations, a deeper understanding of disc viscosity and stellar feedback is needed to understand the details of the accretion strength at periastron and the exact timing of the bursts.

\section{Limitations and caveats}
\label{sec:caveats}

\subsection{Numerical resolution}

The level of refinement used in our simulations does not resolve the regions closest to the actual protostars and this limits the ability to resolve circumstellar discs at small binary separations. In our work, the resolution on the highest level of refinement corresponds to a cell size of $\Delta x \sim$1.95$\au$. \cite{federrath_modeling_2014} find that to have fully converged outflow efficiencies for a simulation box size the same as that used in our work requires cell sizes of $\Delta x \sim 0.06\au$ to resolve the jet launching region. Running simulations with this level of resolution is very computationally intensive and impractical at the moment.

The binary-star systems that form in our simulations have final separations of \mbox{$9$--$10\au$}, which is resolved over \mbox{$\sim4$--$5$} cells. Following the prescription of \cite{artymowicz_dynamics_1994}, if the sink particles host circumstellar discs, they would have a radius of \mbox{$4$--$5\au$}. These discs would only be resolved over approximately \mbox{$2$--$3$} cells. However, since we are studying the accretion behaviour over the entirety of the eccentricity evolution, the circumstellar discs are resolved with more than that for most of the eccentricities, but analysis of perturbations within the circumstellar discs is not possible. We conduct a resolution study in \Cref{sec:appendix}. This study addresses some of the concerns presented here.

\subsection{Radiation effects}

Our simulations do not explicitly calculate radiative transfer. However, our equation of state accounts for some of the radiative effects on the local cell scale (see \Cref{ssec:flash}). There have been works considering both radiative feedback and ideal MHD, mostly concerning cluster formation (\citealt{offner_effects_2009, price_inefficient_2009, myers_fragmentation_2013, myers_star_2014, krumholz_what_2016}). \citet{bate_importance_2009} and \citet{offner_effects_2009} find that radiation feedback suppresses fragmentation of discs, however, this assumes continuous accretion. \citet{stamatellos_episodic_2012} investigate episodic accretion on disc fragmentation and find that discs can still be susceptible to gravitational instability provided that the time between bursts is longer than the dynamical timescale for the growth of gravitational instabilities. \citet{bate_stellar_2012} conclude that the main physical processes involved in determining the properties of multiple stellar systems are gravity and gas dynamics.

Work investigating radiation feedback on star formation in single cores have mostly focused on massive star formation. Because massive stars have higher luminosities, radiation feedback plays a significant role in their formation and evolution. However, \citet{tanaka_impact_2018} found in one- dimensional models of massive star formation that radiation feedback only made a minor contribution to the star formation efficiency, and magneto-centrifugally driven outflows are the dominant feedback process. Three-dimensional radiation hydrodynamic simulations by \citet{klein_feedback_2010} find similar results, with outflows suppressing the effects of radiation pressure and thereby reducing radiation feedback. Despite radiation feedback not playing a dominant role in suppressing accretion, \citet{rosen_massive-star_2019} find that radiation feedback coupled with outflows produces lower accretion rates than just radiation alone onto massive stars.

Concerning radiation feedback in young binaries, \citet{young_binary_2015} carried out 2D SPH simulations of accretion in binaries, varying mass ratio and gas temperature. They found that higher gas temperature resulted in a higher accretion rate onto the primary component. They attribute this to the increased gas sound speed, leading to higher gas viscosity and lowering the viscous timescale. 

It is not clear what the effect of radiative feedback would have on a low mass binary like those produced in our simulations, but the influence of radiative feedback on episodic accretion in binary-star systems should be investigated in future studies.

\subsection{Non-ideal MHD effects}

The non-ideal effects of Ohmic resistivity, the Hall effect and ambipolar diffusion are important on scales $\sim$1.5, $2$--$3$ and $\geq 3$ scale heights, respectively (\citealt{wardle_magnetic_2007, salmeron_magnetorotational_2008, konigl_effects_2011, tomida_radiation_2015, marchand_chemical_2016}). Further away from the disc, the surface layers of discs are expected to be ionised by stellar radiation in the FUV and the ideal MHD limit is a reasonable approximation \citep{perez-becker_surface_2011, nolan_centrifugally_2017}.

Viscosity is an important property dictating the timescale of a circumstellar disc disruption leading to an accretion event. Viscosity in discs has often been attributed to magneto-rotational instability \citep[MRI][]{balbus_powerful_1991} which arises from a combination of Keplerian shearing and magnetic tension. The degree to which MRI is effective is dependent on the level of magnetisation, and because non-ideal MHD reduces coupling between the gas and the magnetic field, it is expected that the effective viscosity would be reduced in non-ideal MHD \citep{ercolano_dispersal_2017}. \citet{zhu_dust_2015} investigated the effect of ambipolar diffusion on MRI via three-dimensional global simulations and found that the MRI was weaker than in the ideal MHD limit, leading to lower viscosity within discs. Thus, non-ideal MHD may reduce the disc viscosity, allowing for a more accurate reproduction of observed accretion behaviour. Non-ideal MHD should be considered in follow-up work when studying accretion discs.

\section{Summary and conclusion}
\label{sec:conclusion}

We ran and analysed MHD simulations of binary-star formation with varying levels of turbulence. We quantified how eccentricity influences the strength of accretion over a binary orbit, what physical mechanism triggers accretion events, and compared the results of our simulations with observational data. We ran two simulations with initial turbulence of Mach~0.1 (\emph{T1}), and Mach~0.2 (\emph{T2}). We find the following main results:

\emph{Dependence of episodic accretion on eccentricity}. We find that orbital phase-correlated episodic accretion occurs in binaries of eccentricity $e>0.2$. For \emph{T1}, we find that episodic accretion peaks for moderate eccentricity ($0.3<e<0.7$). \emph{T2} shows a general linear trend of weak episodic accretion at low eccentricity to strong accretion at high eccentricity. These varying results imply that eccentricity alone does not determine the strength of episodic accretion. We also find that for the highest eccentricities, the strength of episodic accretion is weaker than at moderate eccentricities. This is likely due to other factors such as more severe circumstellar disc disruption at periastron when the eccentricity is extremely high and higher quiescent accretion rates.

\emph{Mechanism triggering accretion events}. Based on high-resolution simulations, we determine that it is primarily torques between the circumstellar disc and the companion that triggers an accretion event. This exchange of angular momentum from the gas to the binary orbit is observed at the beginning of an accretion event, exciting a spiral density wave, which enhances accretion onto the stars. Observations of spiral arms in protoplanetary discs show similar structures to those seen in our simulations \citep{rosotti_spiral_2019}.

\emph{Comparison with observations}. Our simulations are able to reproduce the timing of episodic accretion found in observations, despite the vastly different orbital periods of the observed and simulated binaries. Our simulations produce normalised accretion rates at the peak of the accretion burst that are about a factor of 1.5 to three higher than those observed, and the peak of accretion occurs slightly earlier in terms of orbital phase (at $\phi\sim0.95$) than those observed ($\phi\sim1$). These differences may stem from the fact that our simulations study a much earlier phase of evolution (class 0/I) than the observations we compared to (class II), however, as discussed in \Cref{ssec:observations}, other simulations of Class II binaries were also not able to reproduce observed accretion behaviour. Therefore, we suggest that a deeper understanding of the effects of disc viscosity and stellar feedback is needed to understand the details of episodic accretion at periastron and the exact timing of the accretion bursts.

Overall we find that episodic accretion can be seen in binaries with eccentricity $>0.2$, and the shape of episodic accretion is in good agreement with that found in observations of short-period binaries. This suggests that the accretion behaviour at periastron for eccentric binaries does not significantly depend on the orbital separation. This can have implications for understanding binary triggers in FU~Orionis or EXor-type outbursts. Given that episodes of high accretion last for a certain duration in phase space, this can translate to very different timescales depending on the period of the binary. Based on our simulations, the burst period spans approximately phases $0.8<\phi<1.1$, which is $\sim30\%$ of an orbital period. For a binary of period $>40\yr$ experiencing episodic accretion triggered by a companion, the burst period would last $>10\yr$, i.e. of the order of the observed timescales of FU~Orionis events \citep{hartmann_fu_1996}. The same argument applies to shorter-period binaries with periods $\gtrsim1\yr$ to produce bursts with timescales similar to EXor outbursts \citep{herbig_1993-1994_2001}. However, FU~Orionis events have burst accretion rates of approximately 100 times greater than their quiescent rate. Our simulations do not produce this, but as shown in our resolution study, our measures of the ratio of accretion rate during burst and quiescent periods has not fully converged, and our binaries are still at a very young embedded stage compared to most observational data currently available. Overall, we suggest that signs of episodic accretion should be observable in long-period binaries.

\section*{Acknowledgements}
We thank the anonymous referee for a timely and constructive report. The authors thank B.~Tofflemire for providing the observational data used in this paper. This project has received funding from the European Union’s Horizon 2020 research and innovation programme under the Marie Sklodowska-Curie grant agreement No. 847523 ‘INTERACTIONS’. R.~K.~would like to thank the Australian Government and the financial support provided by the Research Training Programme Domestic Scholarship. The research leading to these results has received funding from the Independent Research Fund Denmark through grant No.~DFF 8021-00350B (T.~H., R.~K.). C.~F.~acknowledges funding provided by the Australian Research Council (Discovery Project DP170100603 and Future Fellowship FT180100495), and the Australia-Germany Joint Research Cooperation Scheme (UA-DAAD). The simulations presented in this work used high performance computing resources provided by the Leibniz Rechenzentrum, the Gauss Centre for Supercomputing (grants~pr32lo, pr48pi and GCS Large-scale project~10391), and the Australian National Computational Infrastructure (grant~ek9) in the framework of the National Computational Merit Allocation Scheme and the ANU Merit Allocation Scheme. The astrophysics HPC facility at the University of Copenhagen, supported by a research grant (VKR023406) from VILLUM FONDEN, was used for carrying out part of the simulations, the analysis, and long-term storage of the results. The simulation software \texttt{FLASH} was in part developed by the DOE-supported Flash Center for Computational Science at the University of Chicago. \texttt{yt} \citep{turk_yt:_2011} was used to help visualise and analyse these simulations.

\bibliographystyle{aa}
\bibliography{references}

\begin{appendix}
\section{Resolution study}
\label{sec:appendix}

To accompany our investigation into the accretion behaviour as a function of eccentricity we conducted a resolution study. Due to computational difficulties, we were only able to run \emph{T2} at a level of refinement of up to 14 (c.f.~\Cref{ssec:simulationsetup}) for enough orbits to carry out this resolution study. Due to this, we assume that the convergence results for \emph{T2} can appropriately be applied to \emph{T1}.

\begin{figure}
	\centerline{\includegraphics[width=1.0\linewidth]{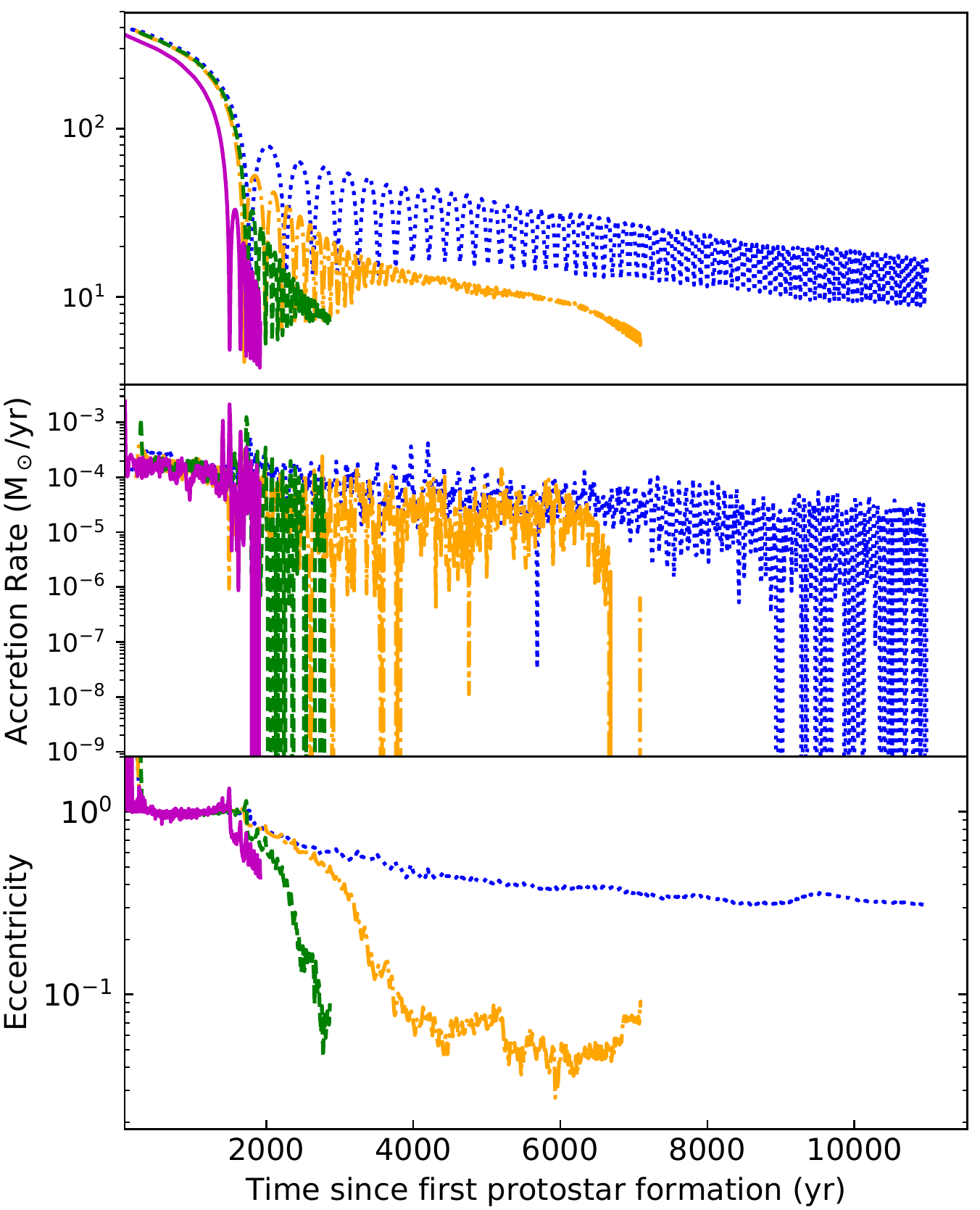}}
	\caption{Same as \Cref{fig:SystemEvolution}, but showing the system evolution for level of refinements of 11 (blue), 12 (orange), 13 (green), and 14 (magenta) for the \emph{T2} simulation setup.}
	\label{fig:system_res}
\end{figure}

\begin{figure}
	\centerline{\includegraphics[width=1.0\linewidth]{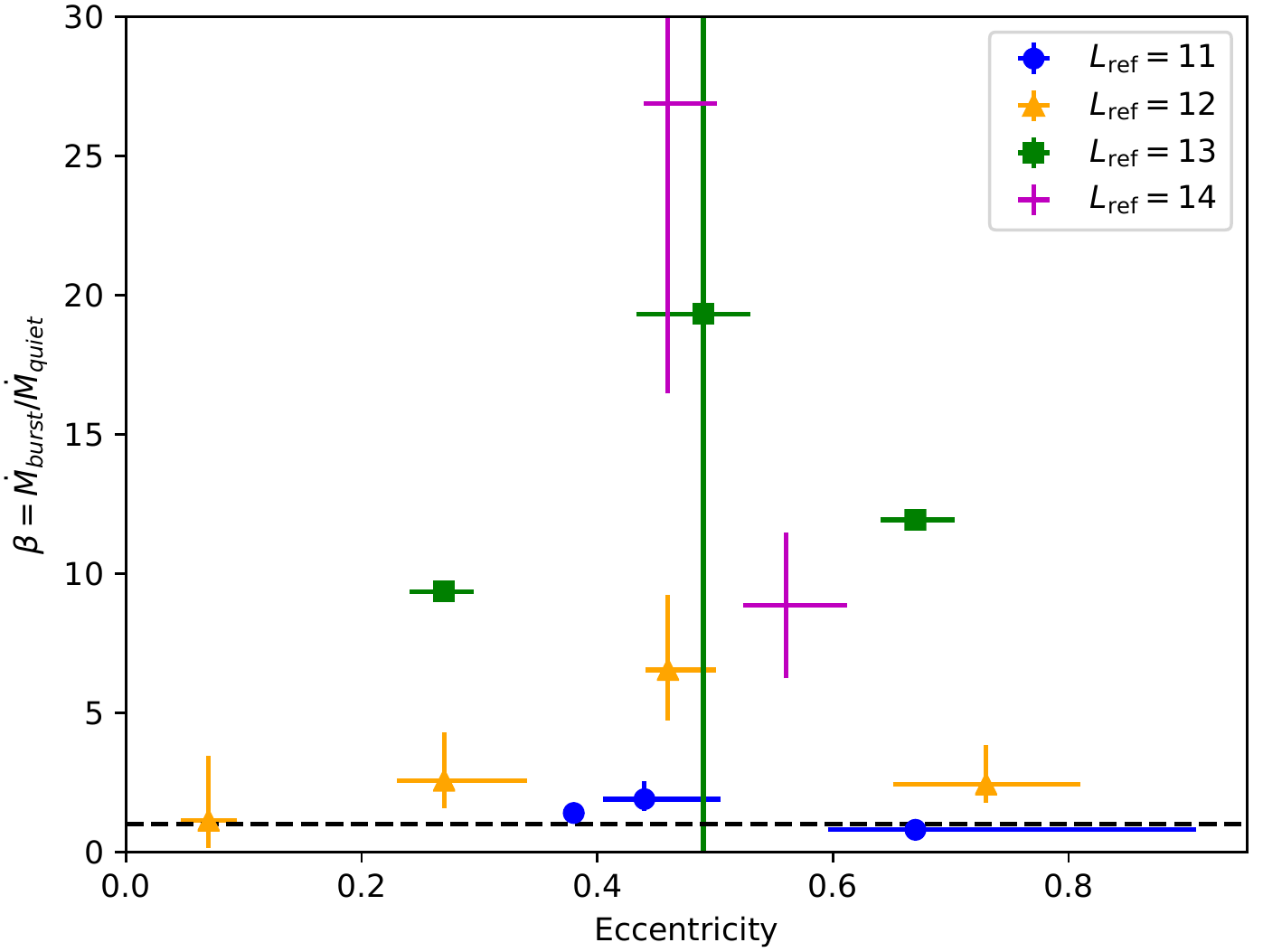}}
	\caption{Same as \Cref{fig:beta}, but showing the values for level of refinements of 11 (blue circles), 12 (orange triangles), 13 (green squares), and 14 (magenta cross) for the \emph{T2} simulation setup.}
	\label{fig:beta_res}
\end{figure}

In \Cref{fig:system_res} we show the separation, total accretion rate, and eccentricity of the simulation at the different resolution levels, similar to \Cref{fig:SystemEvolution}. We see that with higher resolution we observe a faster in-spiral rate, and this has not yet converged with the resolutions tested. Our interpretation is that this is related to resolving the torques in the circumstellar discs. An even higher resolution would be required to properly account for the transfer of angular momentum, in particular in the early phase, where the tidal forces in the discs related to the high eccentricity are the largest. The accretion rate shows a steady decline as the systems evolve and there is no significant variation with resolution. We also observe that the separation of the binary at the first periastron is converged at $\sim$5$\,\mathrm{AU}$ across all resolutions and we take this to be the true closest approach that the binary experiences at the beginning of in-spiral phase.

\begin{figure*}
	\centerline{\includegraphics[width=1.0\linewidth]{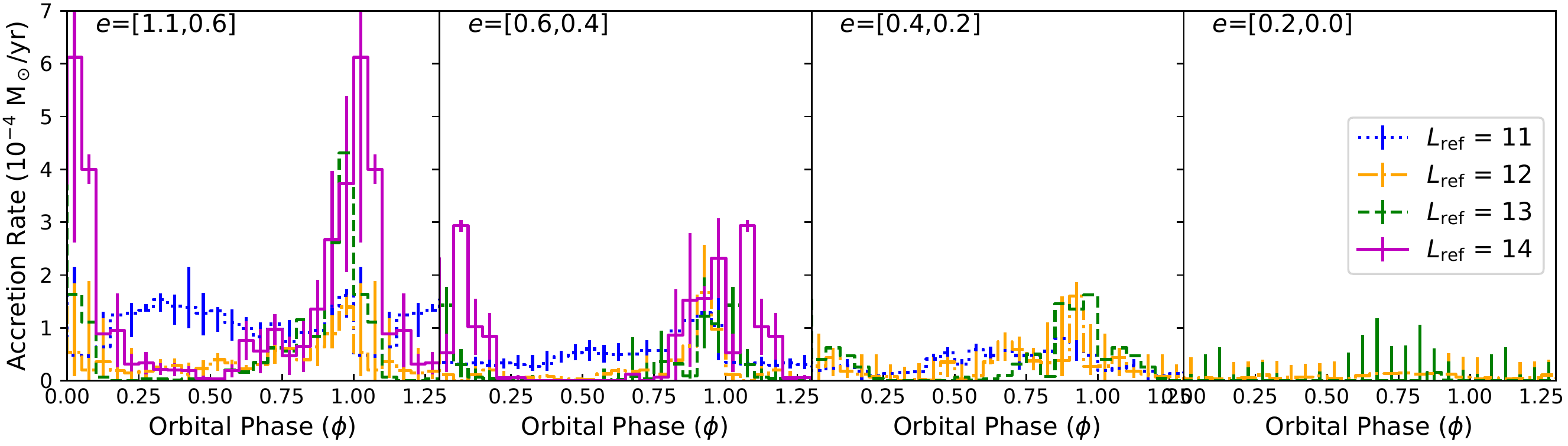}}
	\caption{Same as \Cref{fig:componentAccretion2}, but showing the phase-folded accretion for level of refinements of 11 (blue dotted), 12 (orange dash-dot), 13 (green dashed), and 14 (magenta solid) for the \emph{T2} simulation setup.}
	\label{fig:phase_res}
\end{figure*}

We then phase-folded the accretion to determine whether the accretion behaviour has converged for different eccentricity bins. Due to the different rate of in-spiral between the simulations we adjusted the eccentricity bins to be [1.1, 0.6, 0.4, 0.2, 0.0]. This ensures that at least two orbits are within a bin without having bins that are too wide in eccentricity space. With these bins we show the phase-folded accretion in \Cref{fig:phase_res}. The error bars are the 16th and 84th percentile, as explained in \Cref{ssec:episodic_accretion}, and require at least three orbits in each bin to be calculated. The number of orbits per bin is summarised in \Cref{tab:resolution_summary}.

From \Cref{fig:phase_res} we see that the maximum accretion rate has converged for moderate to low eccentricities, with the accretion rate of the high-resolution simulations being in agreement with the $L_{ref}=12$ simulation mostly analysed in this study. This is shown by the peak accretion of $L_\mathrm{ref}=12$ and $L_\mathrm{ref}=13$ being within the error bars for eccentricities $e\lesssim0.6$. There is some minor variation in the location of the peak in phase-space, with the peak accretion occurring at slightly later phases. This may account for the discrepancy in the location of the simulated and observed peak accretion discussed in \Cref{ssec:observations}.

The resolution study also demonstrates that the maximum accretion rate is in agreement, within the error bars, for high eccentricities at the two highest levels of resolution, but is higher by a factor of 2--3 compared to the $L_{ref}=12$ simulation. We should be wary of this when looking at the accretion behaviour at high eccentricities. An even higher resolution would be required to firmly demonstrate convergence of the full accretion history.

\begin{table}
	\caption{Orbits used in each eccentricity bin for each simulation.}
	\label{tab:resolution_summary}
	\centering
	\begin{tabular}{lcccc}
		\hline
		Bins & [1.1,0.6] & [0.6,0.4] & [0.4, 0.2] & [0.2, 0.0]\\
		\hline
		$L_\mathrm{ref}=11$ & 4 & 12 & 3 & N/A \\
		$L_\mathrm{ref}=12$ & 5 & 4 & 4 & 73 \\
		$L_\mathrm{ref}=13$ & 2 & 4 & 2 & 6 \\
		$L_\mathrm{ref}=14$ & 3 & 3 & N/A & N/A \\
		\hline
	\end{tabular}
\end{table}

Over all eccentricities, higher resolution leads to lower quiescent accretion rates, which results in higher values of $\beta$ (c.f.~Equation 8). The various $\beta$-values are shown in \Cref{fig:beta_res}. The $\beta$-values for $L_\mathrm{ref}=12$ are approximately three times larger than the $\beta$-values for similar eccentricities at $L_\mathrm{ref}=11$. The $\beta$-values for $L_\mathrm{ref}=13$ are approximately three times larger than the $\beta$-values for similar eccentricities at $L_\mathrm{ref}=12$. The $\beta$-values measured from $L_{ref}=14$ are, somewhat, in agreement with that measured from $L_{ref}=12$ and $13$, making it difficult to determine whether this is converged. $\beta$-values greater than 100 would enter the realm of FU-Orionis type outbursts, which fully converged simulations may be able to comment on.

Overall, while the $L_\mathrm{ref}=12$ simulations that are primarily analysed in this paper have not fully converged in all aspects, they have converged concerning the maximum accretion for low to moderate eccentricities. The in-spiral rate of the $L_\mathrm{ref}=12$ simulation is also slow enough that it is possible to study the accretion behaviour as a function of eccentricity without needing large eccentricity bins.

\end{appendix}
\end{document}